\documentclass[prl,twocolumn,showpacs,superscriptaddress]{revtex4}
\usepackage[dvips]{graphicx}
\usepackage{times,epsfig,amssymb}

\begin{document}

\title{Self-organized Model for Modular Complex Networks : Division and Independence}
\author{D.-H. Kim}
\affiliation{School of Physics and Center for Theoretical Physics,
Seoul National University, Seoul 151-747, Korea}
\author{G. J. Rodgers}
\affiliation{Department of Mathematical Sciences, Brunel
University, Uxbridge, Middlesex, UB8 3PH, United Kingdom}
\author{B. Kahng}
\author{D. Kim}
\affiliation{School of Physics and Center for Theoretical Physics,
Seoul National University, Seoul 151-747, Korea}
\date{\today}

\begin{abstract}
We introduce a minimal network model which generates a modular
structure in a self-organized way. To this end, we modify the
Barab\'asi-Albert model into the one evolving under the principle
of division and independence as well as growth and preferential
attachment (PA). A newly added vertex chooses one of the modules
composed of existing vertices, and attaches edges to vertices
belonging to that module following the PA rule. When the module
size reaches a proper size, the module is divided into two, and a
new module is created. The karate club network studied by Zachary
is a prototypical example. We find that the model can reproduce
successfully the behavior of the hierarchical clustering
coefficient of a vertex with degree $k$, $C(k)$, in good agreement
with empirical measurements of real world networks.
\end{abstract}


\maketitle

Recently, considerable effort has been made to understand complex
systems in terms of random graphs, consisting of vertices and
edges~\cite{Strogatz01,Albert02,Dorogovtsev02,Newman03a}. Such
complex networks exhibit many interesting emerging patterns as
follows: First, the degree distribution follows a power-law, $P(k)
\sim k^{-\gamma}$, where the degree is the number of edges
connecting to a given vertex~\cite{Barabasi99}. Such networks,
called scale-free (SF), are ubiquitous in the real world. To
illustrate such SF behavior in the degree distribution, Barab\'asi
and Albert (BA)~\cite{Barabasi99} introduced an {\it in silico}
model: Initially, fully-connected $m_0$ vertices exist in a
system. At each time step, a vertex is newly added and connects to
$m$ existing vertices, which are chosen with a probability
linearly proportional to the degree of target vertex. Such a
selection rule is called the preferential attachment (PA) rule.

Secondly, many real world networks have modular structures within
them. Modular structures form geographically in the Internet
\cite{Eriksen03}, functionally in metabolic \cite{Ravasz02} or
protein interaction networks \cite{Rives03}, or following social
activities in social networks \cite{Girvan02,Guimera02}. In these
modular complex networks, the hierarchical clustering coefficient
of a vertex with degree $k$, denoted by $C(k)$, behaves as $C(k)
\sim k^{-\beta}$ \cite{Ravasz02,Ravasz03}, where the clustering
coefficient is defined as the ratio of the number of triangles
connected to a given vertex to the number of triples centered on
that vertex. Also the clustering coefficient averaged over all
vertices is independent of system size $N$. In the BA model,
however, $C(k)$ is independent of $k$, but depends on
$N$~\cite{Albert02,Ravasz03}, because the BA model does not
contain modules. To understand the behavior of $C(k)$, a
deterministic hierarchical model was introduced by Ravasz and
Barab\'asi ~\cite{Ravasz03}, in which $C(k) \sim k^{-1}$ and the
clustering coefficient $C$ is independent of $N$~\cite{Noh03}.
While it is important to understand the mechanism for the
formation of such modular structure through an {\it in silico}
model, few models have been studied, and none in which the modules
were generated in a self-organized way. Thus it is our goal of
this paper to introduce such a model.

\begin{figure}[b]
\includegraphics[scale=0.18]{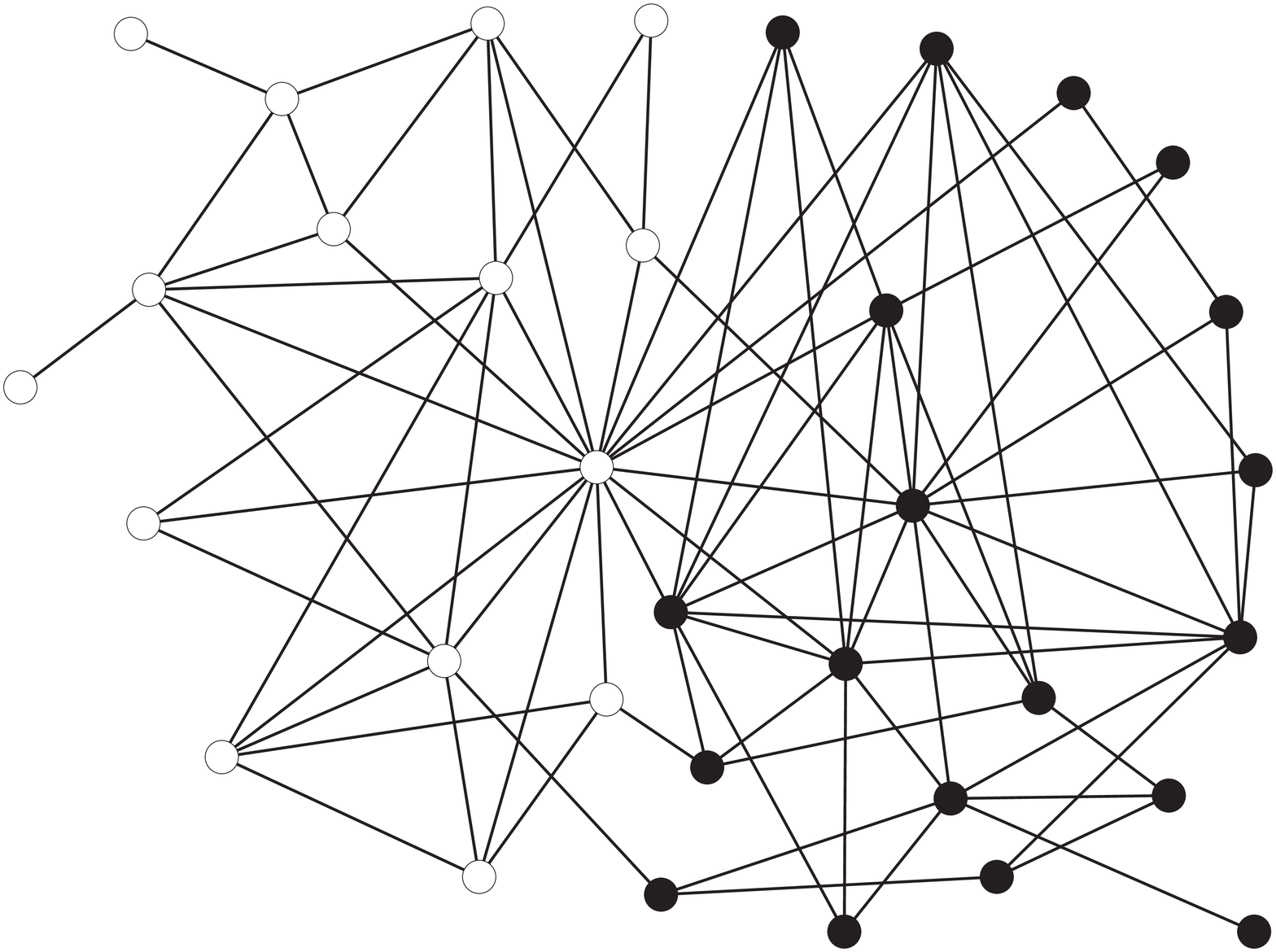}
\caption{A snapshot of the model network with parameters $N=34$,
$m_{0}=4$ and $n=17$, looking similar to the Karate club network
proposed by Zachary. Here two groups are identified by ($\circ$)
and ($\bullet$).} \label{FIG1}
\end{figure}

Thirdly, the degree-degree correlation in real world networks is
nontrivial. The nontrivial behavior is measured in terms of the
mixing coefficient $r$~\cite{Newman02}, a Pearson correlation
coefficient between the remaining degrees of the two vertices on
each side of an edge, where the remaining degree means the degree
of that vertex minus one. Complex networks can be classified
according to the mixing coefficient $r$ into three types, having
$r < 0$, $r\approx 0$, and $r>0$, called the dissortative, the
neutral, and the assortative network,
respectively~\cite{Newman02}. An assortative or dissorative
network can also be identified by a quantity, denoted by $\langle
k_{\rm nn} \rangle (k)$, the average degree of a neighboring
vertex of a vertex with degree $k$ ~\cite{Pastor-Satorras01}. For
the assortative (dissortative) network, $\langle k_{\rm nn}
\rangle (k)$ increases (decreases) with increasing $k$, i.e., a
power law $\langle k_{\rm nn} \rangle (k) \sim k^{-\nu}$ is
satisfied where $\nu$ is negative (positive) for the assortative
(dissortative) network ~\cite{Pastor-Satorras01}.

In this paper, we are interested in modelling modular complex
networks, in particular, forming in a self-organized way. In
social networks, modules represent the communities each individual
belongs to, which may evolve as time passes. The karate club (KC)
network, originally proposed by Zachary~\cite{Zachary77}, is an
example of a social network containing community structures.
Recently, Newman and Girvan ~\cite{Girvan02} studied the KC
network to test a new algorithm for clustering
communities~\cite{Girvan02,Zhou03}. Here we notice that the KC
network contains an important ingredient, division and
independence, needed for the formation of modular structure, in
addition to growth and PA principles as noticed in the BA model.
Thus we introduce a network model evolving by such principles, and
perform numerical simulations for large system size. Indeed, we
find that the model exhibits a characteristic feature of modular
structure, $C(k)\sim k^{-1}$ as much as those for empirical data.

To be specific, the main dynamic process of the evolution of the
KC network is as follows. In a KC, a disagreement develops between
the administrator of the club and the club's instructor as time
goes on, ultimately resulting in the instructor leaving (division)
and founding a new club (independence), accompanied by about half
the original club's members. This generic feature of division and
independence can be observed in many other social communities such
as schools, companies, churches, clubs, parties, etc. For example,
in the coauthorship network, a graduate student publishes papers
with her/his thesis advisor, so that they are connected in a
coauthorship network. When she/he graduates and becomes a
professor in another school (division), she/he also get her/his
own students, creating a new group (independence).

To model the evolution of the KC network, we modify the BA model
by assigning a color to each vertex. The color assigned to each
vertex indicates the group the vertex belongs to. The dynamic rule
of our model is as follows:

(i) \textbf{BA model (Growth and PA)} : Initially, there exist
$m_0$ vertices. They are fully connected. Each vertex $i$ is
assigned the same index of color $\mu_{i}=1$. Thus the total
number of distinct colors $q=1$. At each time step, a vertex is
introduced and connects to $m$ existing vertices following the PA
rule. Here $m$ is not fixed, but is distributed uniformly among
integers in the range $[1,m_0]$. The new vertex $j$ is also
assigned the index of color $\mu_{j}=1$ and this process is
repeated until the number of vertices reaches $n$, a cutoff of the
group size. This process defines the first group $q=1$.

(ii) \textbf{Division and independence} : Then we identify the two
vertices $i$ and $j$ among the group $q$ with the largest and the
second largest degree, respectively, for division and
independence. Then the vertex $j$ declares independence and
changes its color to a new one, $i.e.$, $\mu_j=q+1$. Then, each
remaining vertex $k(\neq i,j)$ in the group having the same color
as vertex $i$ measures the distances $d(k,i)$ and $d(k,j)$ to the
vertices $i$ and $j$, respectively. If $d(k,i) \leq d(k,j)$, then
the vertex $k$ keeps the index of color as it is, otherwise, it
changes its index of color to that of $j$. Then the system
comprises of $q+1$ different groups, and then $q+1 \rightarrow q$,
by definition. So the newest group has the new color $q$.

(iii) \textbf{Growth and PA again} : If $q>1$, then a newly added
vertex $\ell$ chooses one of $q$ colors, say $\mu_{\ell}$, with
equal probability, and $m$, the number of outgoing links, also
randomly from the integers $1,\ldots,m_0$. Then $m$ existing
vertices are chosen in the group with the color $\mu_{\ell}$
following the PA rule, and $m$ edges are inserted between them and
the new node. This process is repeated until the number of
vertices of any group reaches $n$ again. After then, we repeat the
step of division and independence (ii) in that group only.

\begin{table}[b]
\begin{tabular}{|c|c|c|c|c|c|}
\hline
Name & $N$ & $\langle k \rangle$ & $d$ & $r$ & $C$\\
\hline
~~Zachary's~~ & ~~34~~ & ~~4.59~~ & ~~2.41~~ & ~~-0.48~~ & ~~0.59~~\\
Ours      & ~~34~~ & ~~4.61~~ & ~~2.54~~ & ~-0.19~(-0.22)~ & ~~0.56~~\\
\hline
\end{tabular}

\caption{Mean degree $\langle k \rangle$, the diameter $d$, the
assortativity coefficient $r$, and the clustering coefficient $C$
obtained from Zachary's KC network and from ours with parameter
$N=34$, $m_{0}=4$ and $n=17$. All the numerical values for the
model are averaged over ten configurations. Note that Zachary
presumed that the edge between the administrator and the
instructor of the club no longer hold upon division and
independence. Following the Zachary's way, we obtain $r=-0.22$ in
our model.}
\end{table}

\begin{figure}
\includegraphics[angle=270, scale=0.3]{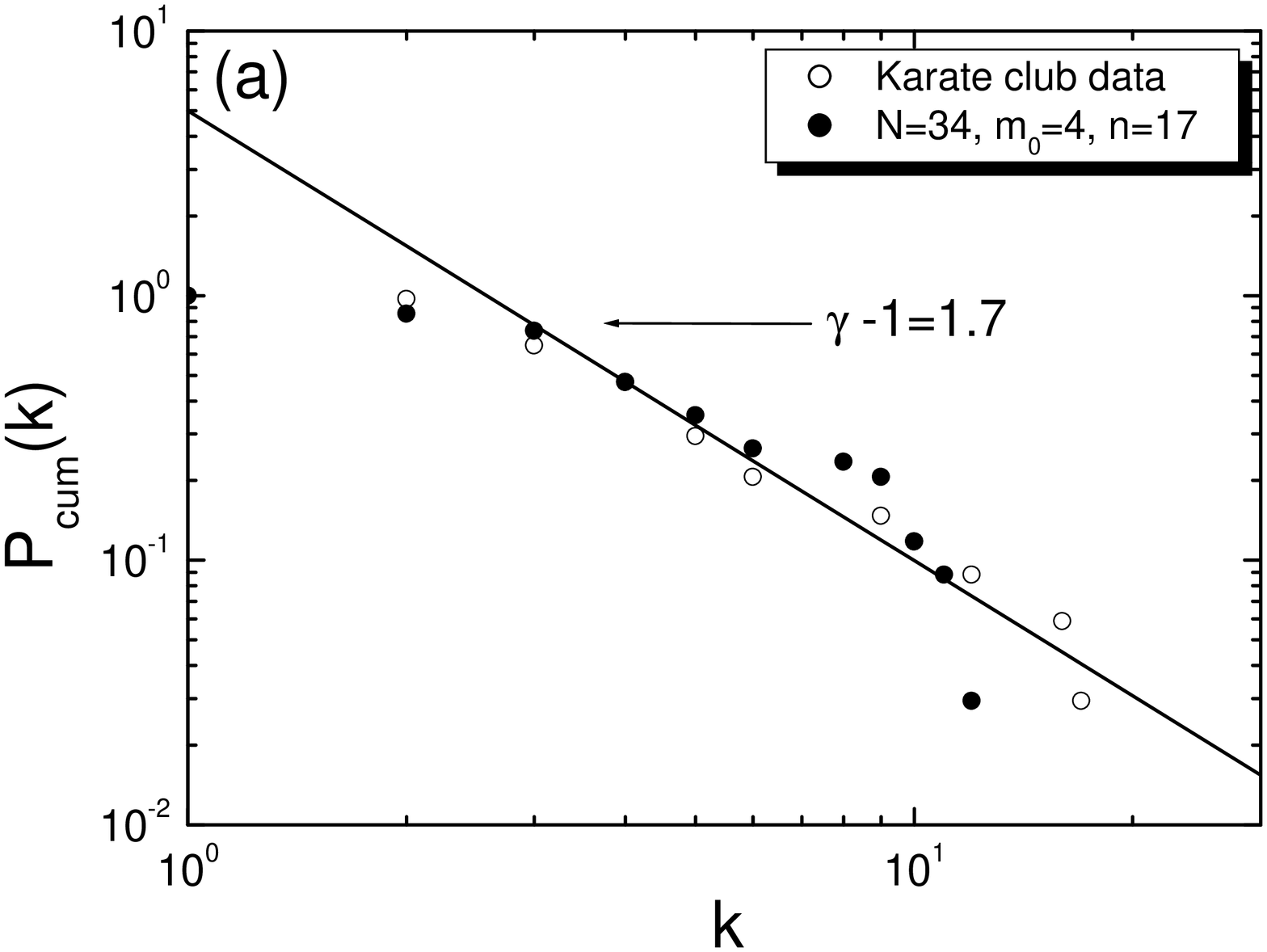}
\includegraphics[angle=270, scale=0.3]{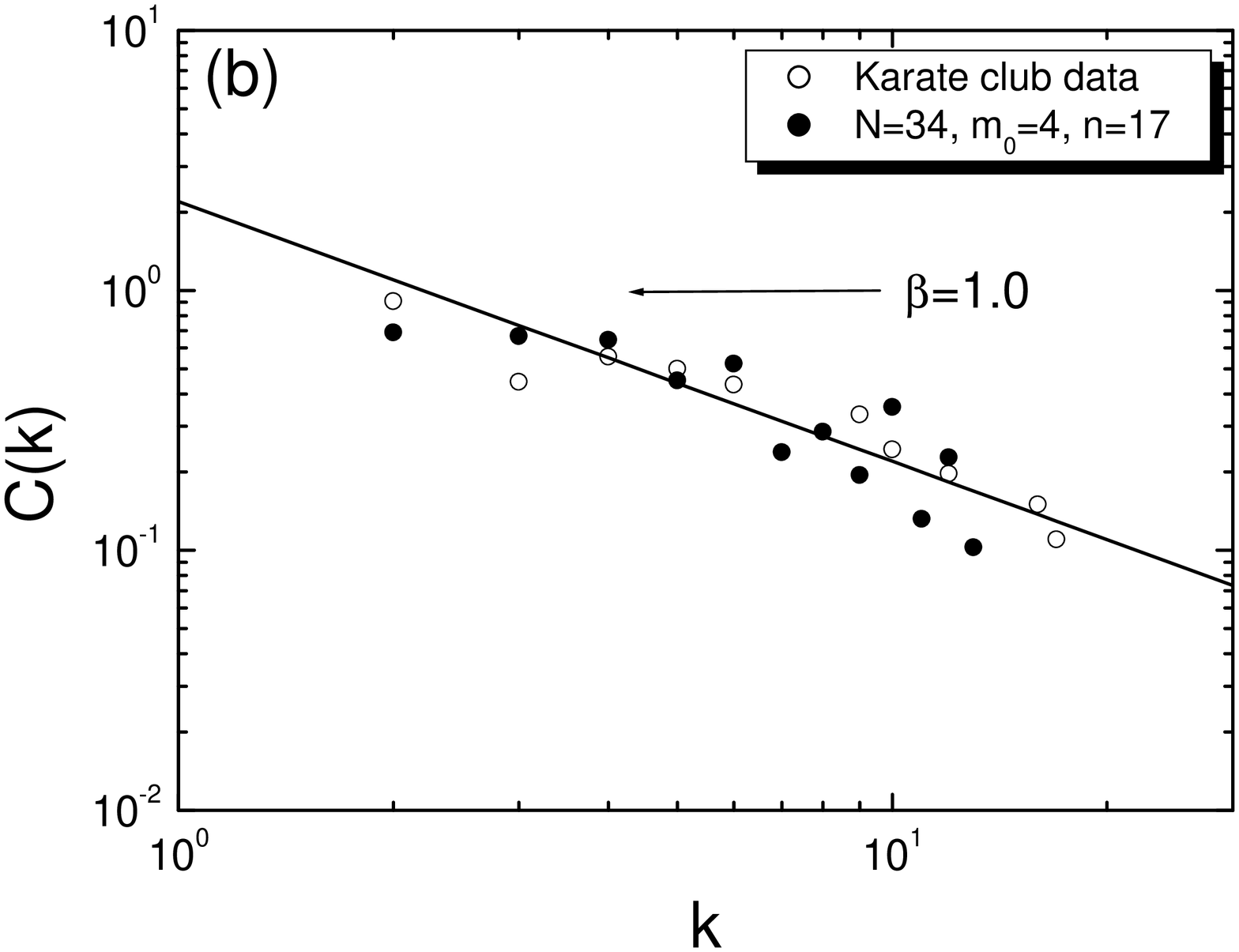}
\includegraphics[angle=270, scale=0.3]{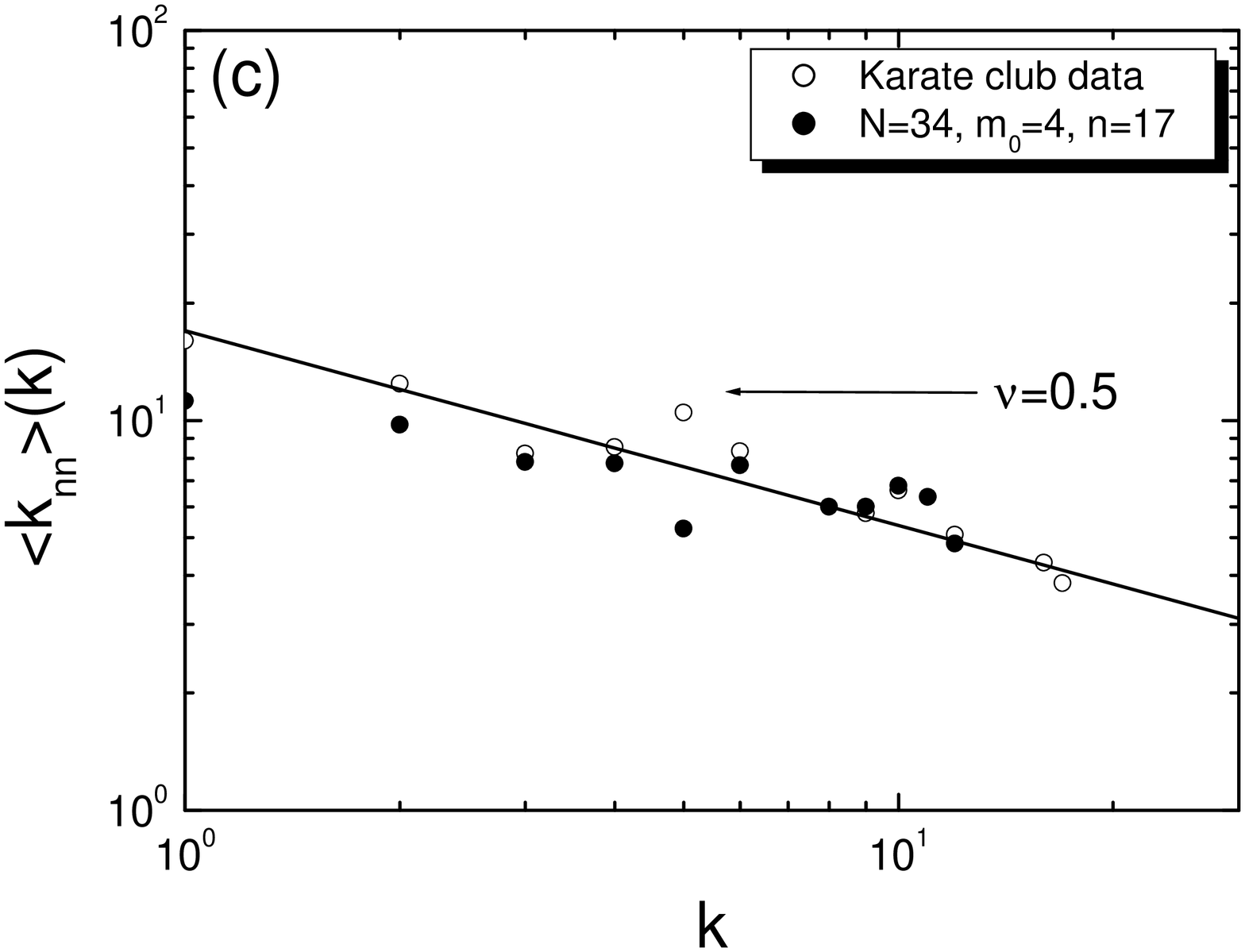}
\caption{Plots of the cumulative degree distribution $P_{\rm
cum}(k)$ (a), the clustering coefficient $C(k)$ (b), and $\langle
k_{\rm nn} \rangle (k)$ (c) versus degree $k$. In all, the
empirical data and the data from the model are denoted by
($\circ$) and ($\bullet$), respectively. The parameters for the
model network are the same as used in FIG.~1. Lines are drawn as a
guide to the eye.} \label{FIG2}
\end{figure}

\begin{figure}
\includegraphics[angle=270, scale=0.3]{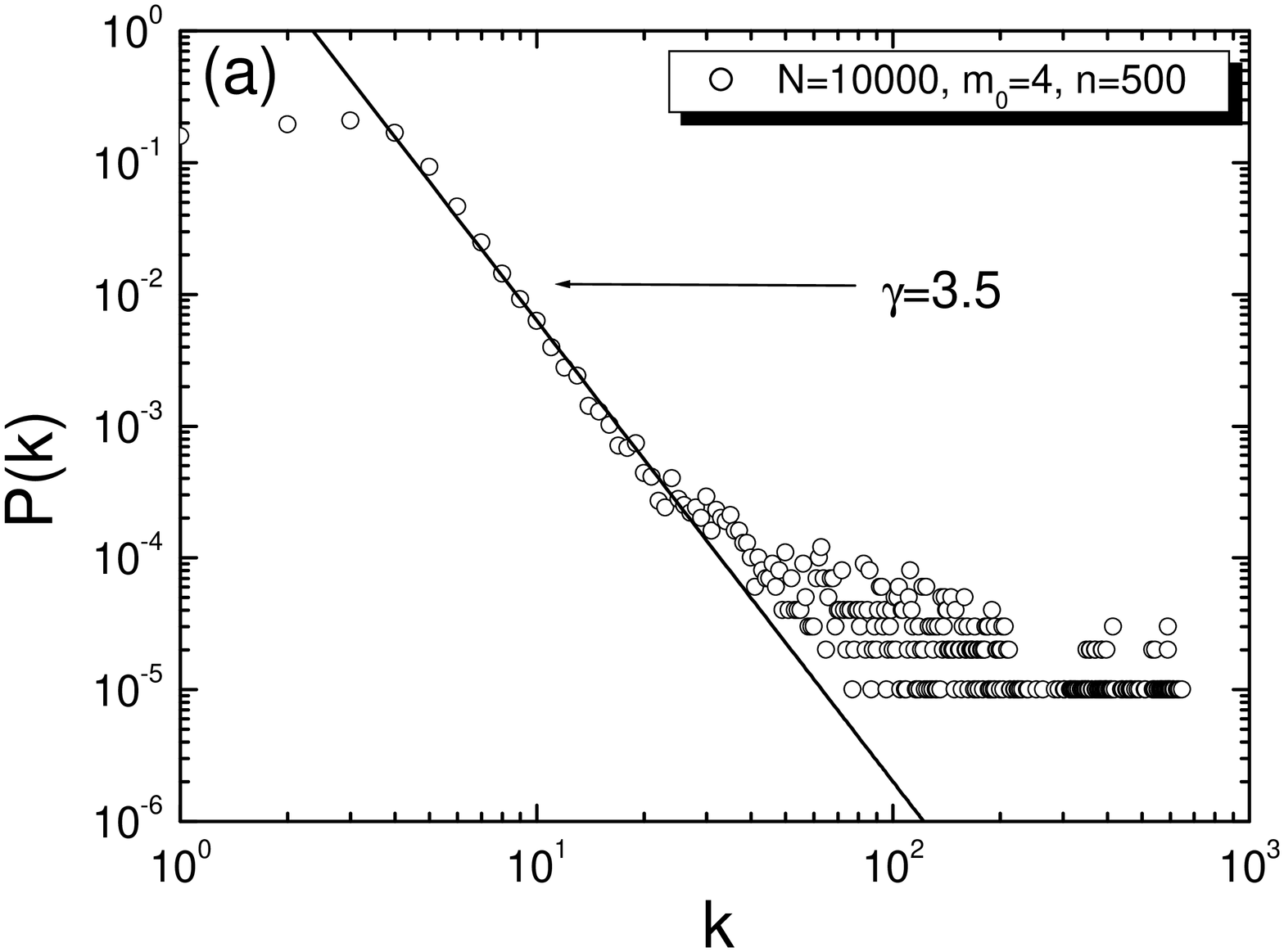}
\includegraphics[angle=270, scale=0.3]{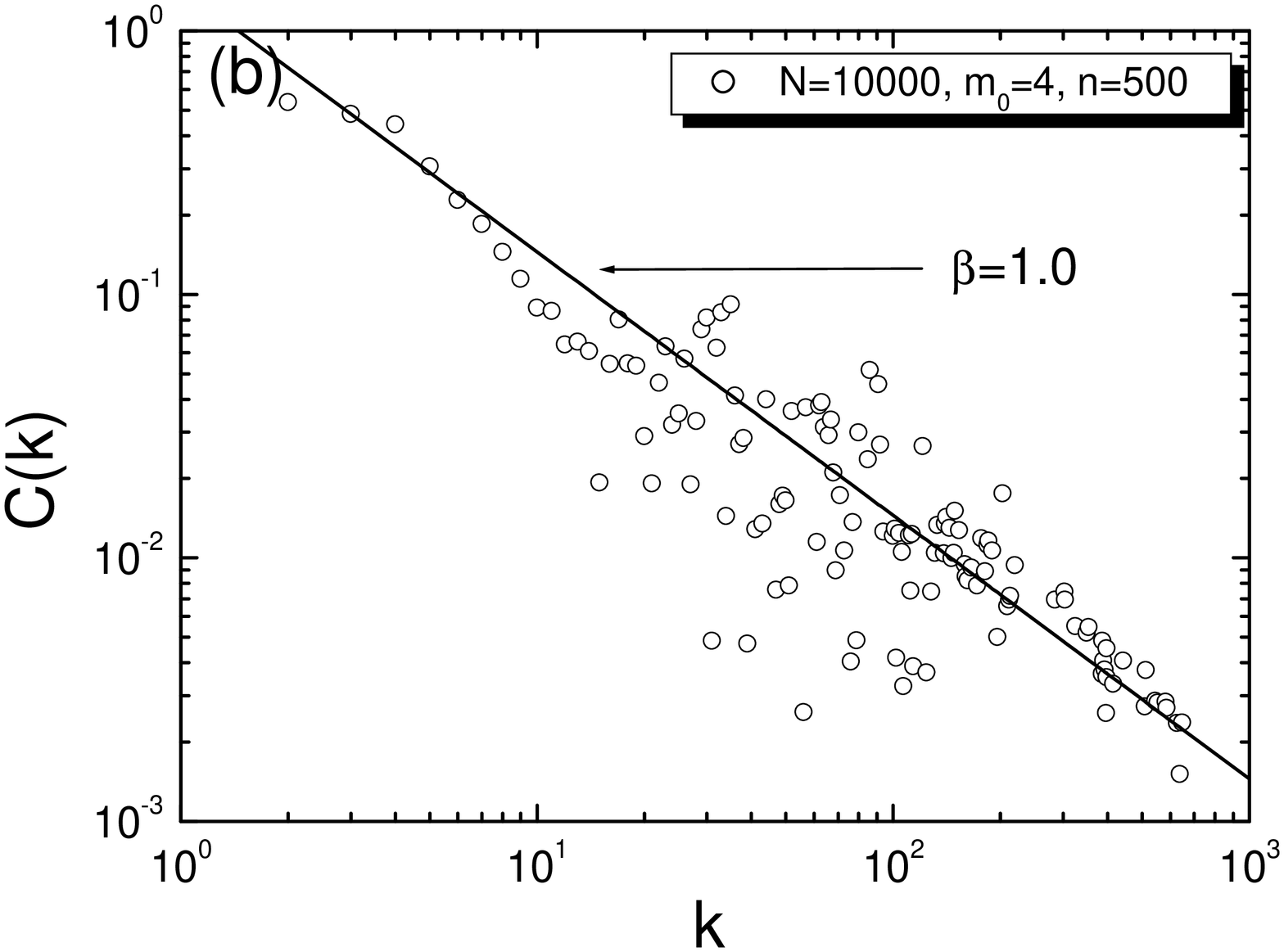}
\includegraphics[angle=270, scale=0.3]{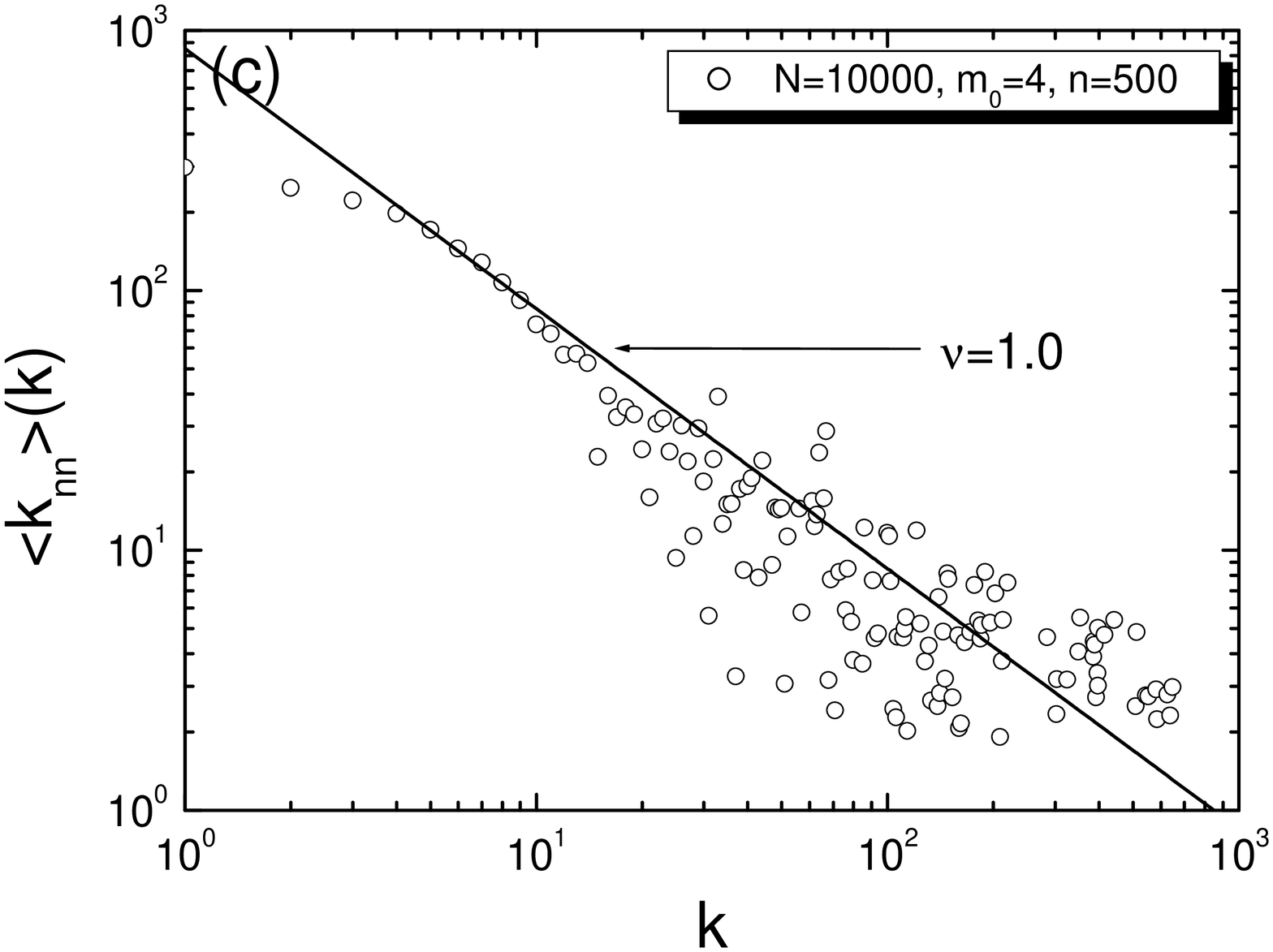}
\caption{(a) Plots of the degree distribution $P(k)$ (a), the
clustering coefficient $C(k)$ (b), and $\langle k_{\rm nn} \rangle
(k)$ (c) versus degree $k$. The data in all figures are obtained
with parameters $N=10000$, $m_{0}=4$ and $n=500$.
In this case we obtain the mean degree $\langle k \rangle = 4.98$,
the diameter $d=4.87$, the assortativity coefficient $r=-0.24$,
and the clustering coefficient $C=0.42$.} \label{FIG3}
\end{figure}

The network constructed in this way is shown in FIG.~1 based on
the same number of vertices as the empirical data of the KC
network. The structure of the model is different from the BA model
due to the presence of modular structure. Note that in our model,
one vertex may transfer from one group to another as time goes on,
that is, a vertex can change its color as it transfers to a new
group. This characteristic is different from that of the
$q$-component static model proposed by the current
authors~\cite{Kim03}, where each individual belongs concurrently
to $q$ different groups such as high school alumni, college
alumni, company, etc. Those two models may reflect different
aspects of our social community.

Based on the empirical data by Zachary, we obtain topological
properties of the KC network, which are listed in TABLE 1 and
FIG.~2. Until now, it has been believed that social networks are
generally assortative ~\cite{Newman02,Newman03b}. But, in
``division and independence" social networks such as the KC
network, each element is connected to the others in a hierarchical
way, without any mediator, leading to a dissortative network, as
shown in TABLE 1 and FIG.~2. Since different colors represent
distinct modules~\cite{Ravasz02,Ravasz03} or communities
\cite{Girvan02}, connections are very tight. Thus it is expected
that the clustering coefficient $C$ is non-trivially large
\cite{Newman03b}. TABLE I shows the dissortativity and the
highly-clustered nature of the KC network and our model.
Agreements between the two are excellent except for the mixing
coefficient $r$.  Note that the $r$ value of the model is not
close to zero although we used the BA-type random attachment rule.
It should be noted that the large value of $C$ is obtained in a
self-organized way. FIG.~2 shows that the degree distribution,
$P(k)\sim k^{-2.7}$, the hierarchical clustering coefficient,
$C(k)\sim k^{-1.0}$, and $\langle k_{\rm nn} \rangle (k) \sim
k^{-0.5}$ of the KC network, which are also in good agreement with
those obtained from the present model network. Such agreements
indicate that our simple model captures the essential topology of
the KC network.

More generally, we investigated the topological properties of our
model network for large $N$ with various $n$. In FIG. 3, we
consider the case of $N=10000$, $m_{0}=4$, and $n=500$. FIG. 3(a)
shows the degree distribution of our model. It seems that $P(k)$
follows a power law with the exponent $\gamma \approx 3.5$, but
that there exists plateau behavior for large $k$. The plateau for
large $k$ is caused by the artificially uniform cutoff of the
group size $n$, which should be modified to fir the empirical
data, if available. If $n$ is not uniform, but is made stochastic
following, for example, a power law, then the shape of the plateau
would change accordingly. FIG. 3(b) shows the hierarchical
clustering coefficient $C(k)$ behaving as $\sim k^{-1.0}$, which
is in good agreement with the Ravasz-Barab\'asi model
\cite{Ravasz03}. FIG. 3(c) shows $\langle k_{\rm nn} \rangle (k)$,
showing a dissortative behavior. The exponent $\nu$ is somewhat
different from the one measured in the small network in FIG.~2(c),
because the size of $N=34$ in FIG. 2 may be too small to measure
the exponent $\nu$, as can be seen in small $k$ of FIG. 3(c). Also
there occurs a plateau region for large $k$ in $\langle k_{\rm nn}
\rangle (k)$. The dissortative behavior ($\nu > 0$) is caused by
hierarchical organization inside a group.

\begin{figure}
\includegraphics[angle=270, scale=0.18]{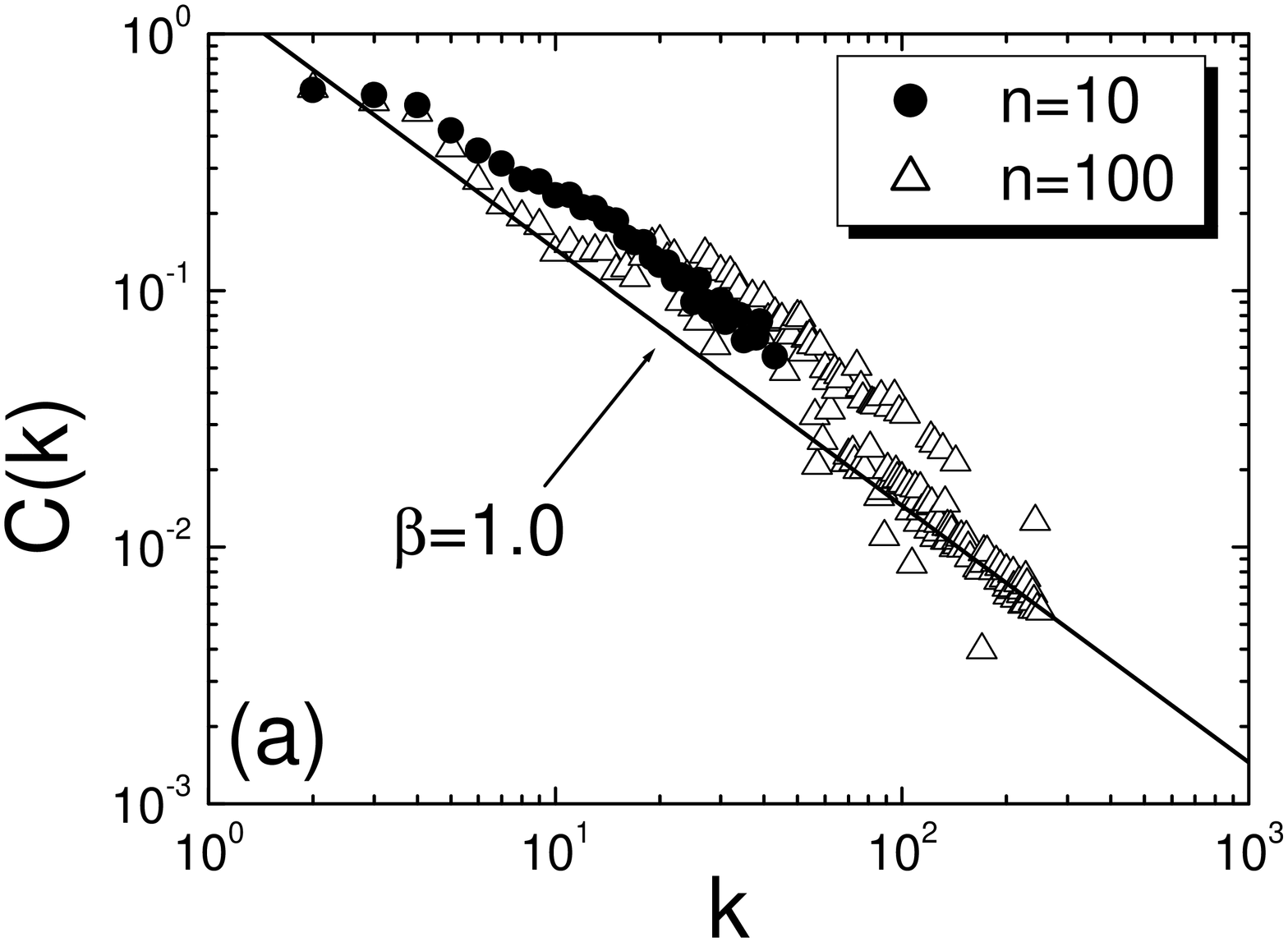}
\includegraphics[angle=270, scale=0.18]{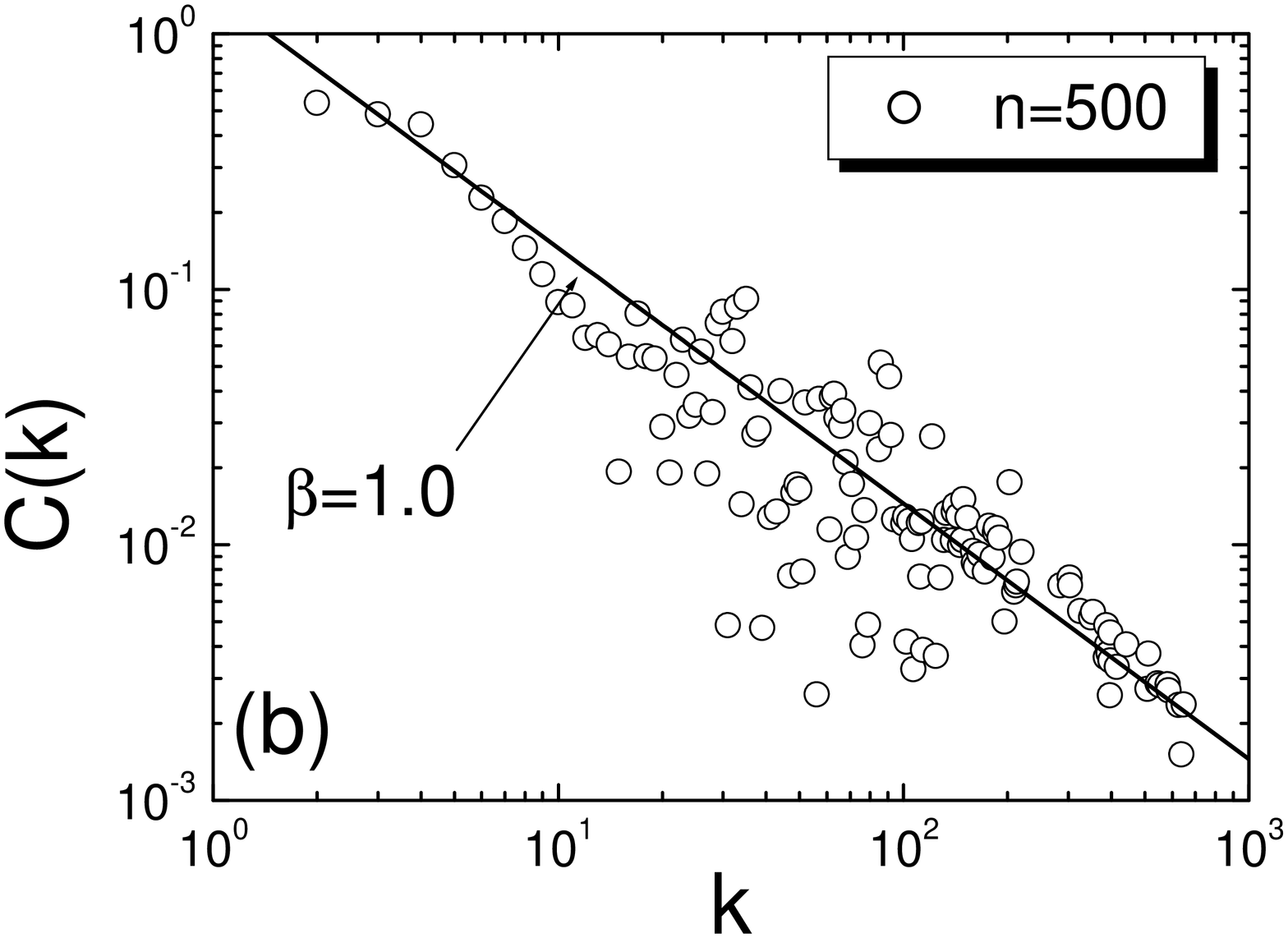}
\includegraphics[angle=270, scale=0.18]{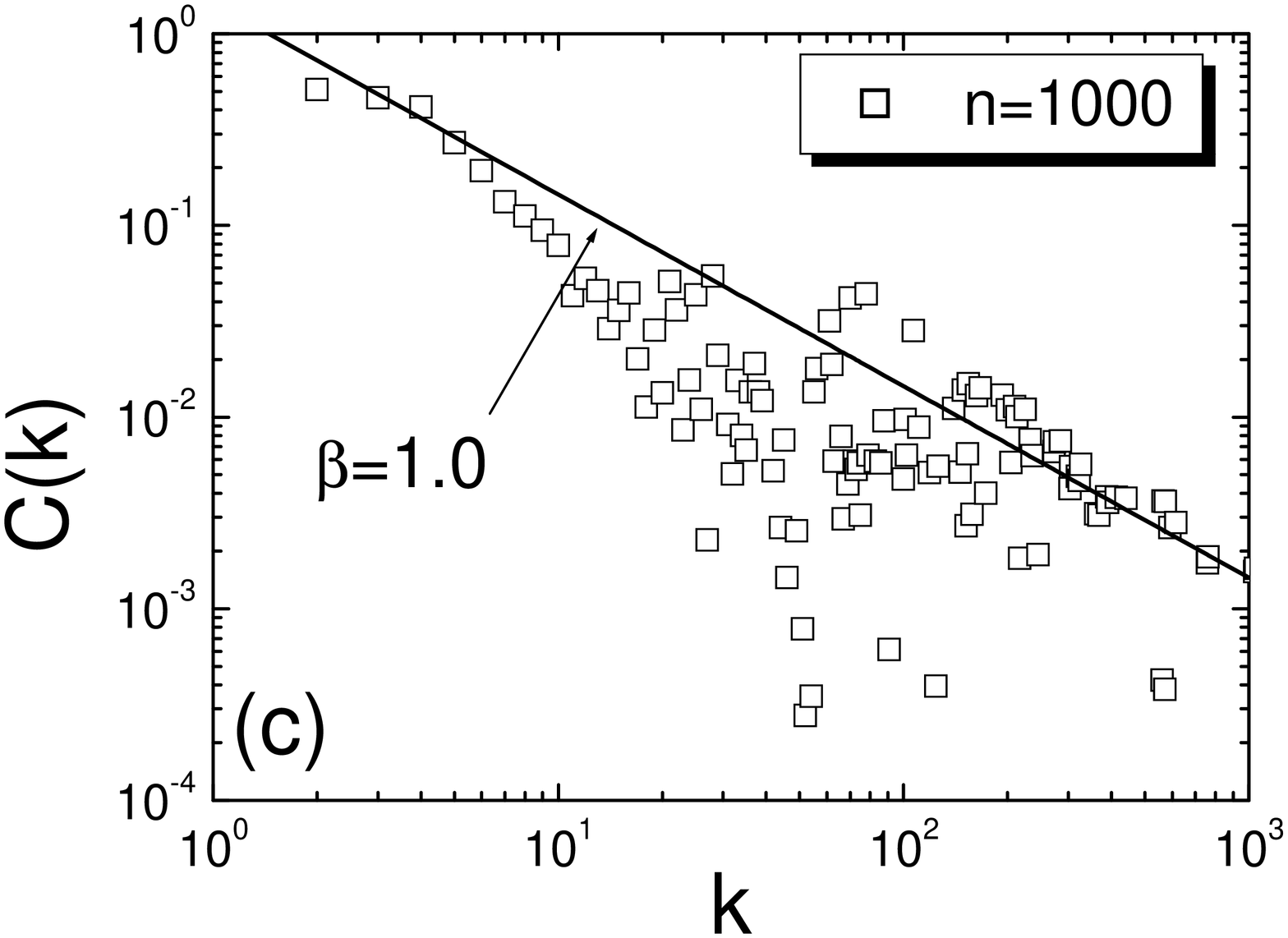}
\includegraphics[angle=270, scale=0.18]{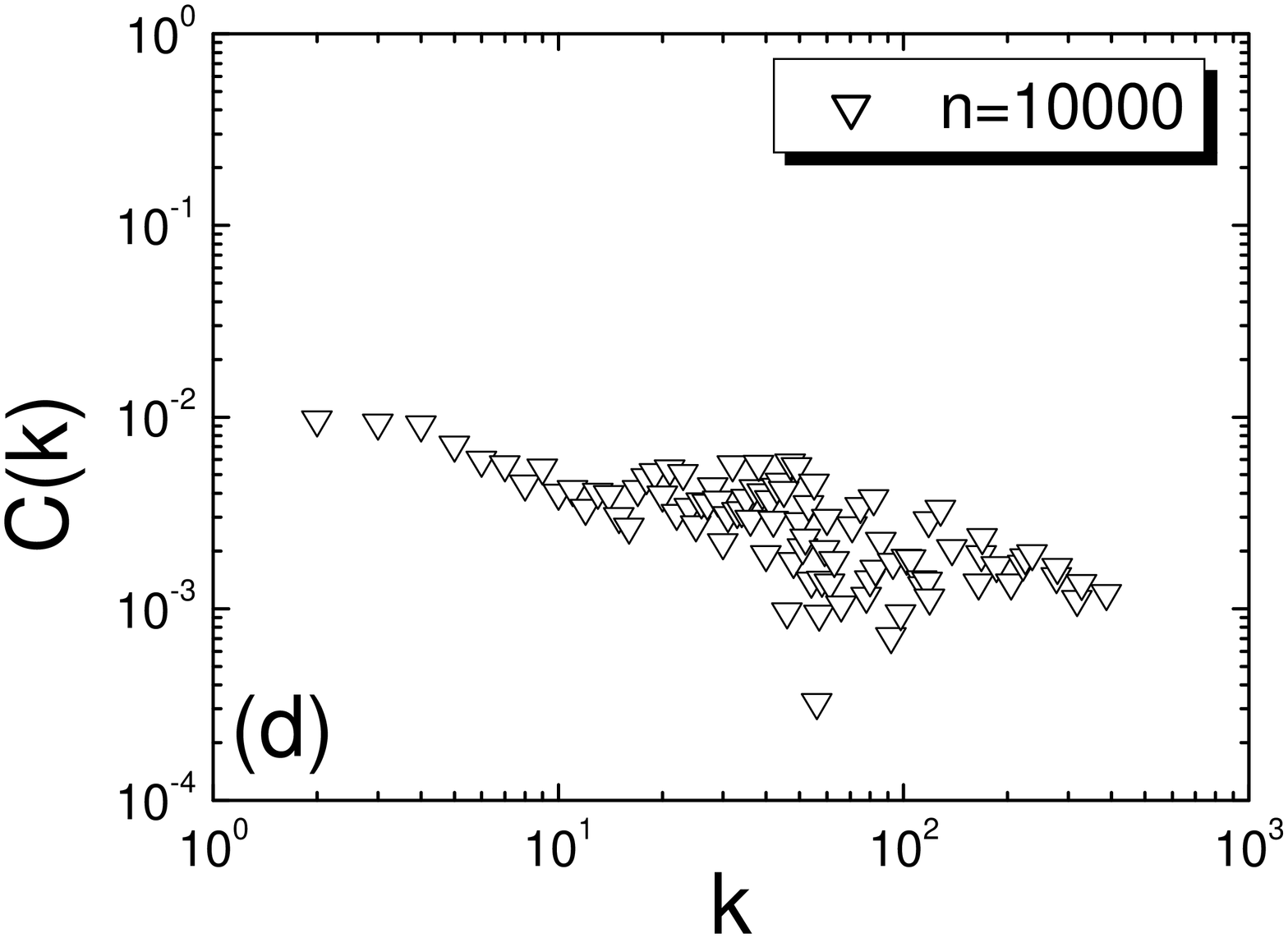}
\caption{The clustering coefficient $C(k)$ versus degree $k$
obtained with the parameters $N=10000$, $m_{0}=4$, and $n=10$ and
100 (a), 500 (b), $n=1000$ (c) and 10000 (d).} \label{FIG4}
\end{figure}

FIG. 4(a) and (b) show the $n$-dependence of the hierarchical
clustering coefficient $C(k)$. When $n$ is very small with respect
to network size $N$, $C(k)$ behaves as $\sim k^{-1.0}$, but as $n$
increases to $N$, $C(k)$ deviates from the power law $C(k) \sim
k^{-1.0}$. The $n=10$ case shows the clear power law behavior. For
$n=100$, a scattered behavior occurs in the middle of the power
law regime. This is found in the actor network (FIG. 3(a) of Ref.
\cite{Ravasz03}). For $n=500$, many points are scattered in the
middle of the power law regime, which is similar to the empirical
results from the Internet autonomous system (FIG. 3(d) of Ref.
\cite{Ravasz03}). For $n=1000$, most of the data points for $C(k)$
are scattered more diversely with $k$, which is similar to the
results from the World Wide Web (WWW) (FIG. 3(c) of Ref.
\cite{Ravasz03}). When the group size $n$ approaches the network
size $N$, $C(k)$ of our model reduces to that of the BA model
(FIG. 2(b) of Ref. \cite{Ravasz03}).

\begin{figure}
\includegraphics[angle=270, scale=0.3]{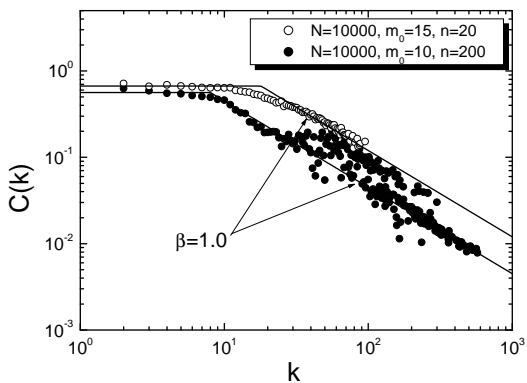}
\caption{The clustering coefficient $C(k)$ versus degree $k$
obtained under two selected conditions with large values of
$m_{0}$.} \label{FIG5}
\end{figure}

FIG. 5 shows the $m_{0}$-dependence of the hierarchical clustering
coefficient $C(k)$. For large $m_{0}$ values, we can see clearly
that $C(k)$ has both a plateau regime from $k=2$ to $k \simeq
m_{0}$ and a power law regime satisfying $C(k) \sim k^{-1.0}$
beyond that degree. When $m_{0}$ approaches the group size $n$,
such as when $m_{0}=15$ and $n=20$, i.e., when vertices inside one
module are nearly fully-connected, such a plateau with a $C$ value
near 1.0 appears. We can thus say that the actor and language
networks (FIG. 3(a) and 3(b) of Ref. \cite{Ravasz03}) have modules
composed of nearly fully-connected vertices. Our model can thus
explain most of the hierarchical clustering structures of real
world networks qualitatively well, when the two parameters $m_{0}$
and $n$ are properly selected. As an example, the case of
$m_{0}=10$ and $n=200$ of FIG. 5 shows a plateau regime as well as
a scattered behavior in the middle of the power law regime, which
are very similar to the actor network (FIG. 3(a) of Ref.
\cite{Ravasz03}).

In conclusion, we have generalized the BA model by assigning a
color to each vertex for the purpose of modelling modular complex
networks in a simple way. The model evolves with time under the
principle of division and independence, in a manner reminiscent of
the KC network. Through this model, we confirmed the behavior of
the hierarchical clustering coefficient, which is in accordance
with the ones obtained from the deterministic hierarchical
structure and the empirical data such as the Internet, the WWW,
and the actor networks \cite{Ravasz03}. Also it was found that our
model exhibits an dissortative mixing behavior as observed in the
KC network. Our model can be modified in various ways, for
example, diversifying the group size cutoff $n$, to fit real world
networks. Finally, we suggest that the principle of division and
independence could be used in constructing modular complex
networks in various fields, for example, bio-complex networks,
where the strong mutation of a gene may correspond to transferring
from one group to another \cite{Goh03}.

\begin{acknowledgments}
This work is supported by the KOSEF Grant No. R14-2002-059-01000-0
in the ABRL program, Korea and by the Royal Society, London.
\end{acknowledgments}

\end{document}